# Comparing different qualitative methods to understand user experience in Saudi Arabia

Aisha Ahmed AlArfaj[1] & Ellis Solaiman[2]
[1] College of Computer and Information Sciences, Princess Nourah Bint Abdulrahman University, Riyadh, Saudi Arabia
[2] School of Computing, Newcastle University, Newcastle Upon Tyne, UK
aiaalarfaj@pnu.edu.sa
ellis.solaiman@newcastle.ac.uk

**Abstract.** The HCI field has seen a growing body of qualitative research, making use of a wide range of activities and methods. Interviews and workshops are some of the main techniques used to help understand user needs and to conduct co-design activities with them. However, these methods might be conducted in various ways and have different advantage and disadvantages. An important aspect influencing the types of activities and methods used is the culture of research participants. This paper aims to compare the research methods conducted in the context of the Saudi Arabian culture. It provides a reflection on the methods used to understand user needs when designing social commerce platforms, including interviews, co-design workshops and critique design workshops. We found that each method has its positives and negatives in terms of user preferences, and can help to obtain useful information at different levels of detail. For example, conducting semi-structured interviews by text was preferred by participants who are at home with their families. However, they can be slower than other methods.

# Introduction

This study employed a combination of a user-centred approach and co-design methodology to develop an in-depth understanding of the current use of social commerce (s-commerce) platforms in Saudi Arabia, aiming to improve user experience and enhance trust in these platforms. The aim of this paper is to present and discuss the methods conducted in this research, highlighting their advantages and disadvantages from practical and cultural perspectives.

A qualitative approach is needed to help researchers study and understand user needs. Within this approach, several methods can be used in human–computer interaction (HCI) research. However, some of these methods present potential issues or challenges, such as high financial cost or being very time-consuming to implement. Moreover, the same methods can be conducted in various ways to obtain different information and details. For example, interviews could be conducted by phone, message or face-to-face. Moreover, they could include activities such as sharing maps and picture cards, or be combined with other methods, such as diary studies, to collect more information on participants' experiences.

For this research, five studies were conducted, employing a mix of methods (interviews, observation, co-design workshops and critique workshops). The first study sought to gain a general understanding of the use of social commerce based on buyer experiences (using text and voice interviews). The second study developed a broad knowledge of the use of social commerce from the sellers' perspective, based on understanding what sellers provide to buyers and how they attract them and gain their trust (using face-to-face & text interviews with online observation). The third study examined sharing activities and compared the use of social commerce and e-commerce (using face-to-face interviews, including drawing a sharing map with a diary). The fourth study employed co-design workshops to finalise the design recommendations (using co-design workshops including screenshots of interfaces and card sorting). Finally, the fifth study aimed to provide a critique of the social commerce platform mock-ups (using critique design workshops, including the mock-ups).

The motivations behind this work are that there is a need to understand the routines and behaviours of s-commerce shoppers to be able to develop or improve social commerce platforms effectively. However, there are few studies that have explored the use of social media to conduct commercial activities (Huang and Benyoucef 2013). In addition, while there is a growing body of literature on s-commerce around the world (Baethge et al. 2016), few studies have been conducted in the context of Saudi Arabia. Most of these studies employing quantitative approaches. Therefore, this novel work was conducted in a culture little explored previously, Saudi Arabia, using qualitative methods.

This paper compares the methods and examines their usefulness. It provides pros and cons for each method. It also explores cultural differences. Finally, it develops guidelines for conducting future research.

# Background

## Research approach

This research focuses on the use of s-commerce in Saudi Arabia. Social commerce can be defined as "the use of Internet-based media to enable users to participate in the selling, buying, comparing, and sharing of information about products and services in online marketplaces and communities" (Busalim et al.

2016). Most of the definitions refer to two main components: social media and e-commerce. According to recent literature (Huang and Benyoucef 2013), there are two main categories of s-commerce:
1. Social media with commercial features enables users to have a personal profile and share information and to search for and buy products using social networking sites (SNSs).
2. Traditional e-commerce websites with social communication and sharing activities, enabling users to rate, review and post comments, as well as to join a community and be able to use social plugin and buy.

Most existing studies in the field of s-commerce have been quantitative in nature (Esmaeili and Hashemi G 2019) and questionnaires have most commonly been used to collect data in Saudi Arabia (Alghamdi et al. 2015)(Nassir and Leong 2017). However, there is a need to gain an in-depth understanding of the current use of s-commerce and how to improve and enhance trust and for such research concerning the use of technologies and the challenges faced, it is recommended to conduct qualitative inquiries to examine the interaction behaviour (Lazar et al. 2017). Whereas quantitative studies address what people do, qualitative studies can help researchers understand why they do what they do (Dimond et al. 2012).

User-centred design (UCD) considers the user as a subject (Sanders and Jan Stappers 2017), which means the focus is on end users (Baxter et al. 2015). The user-centred approach can help to gather information and understand users' experiences through a variety of methods (Baxter et al. 2015). The development of the research design is based on the users' tasks and goals, investigating the context and behaviour of users and considering the users' characteristics (Preece et al. 2015). In HCI, several methods can be employed to study the use of an interface or application, the most common of which are observation, field studies, surveys (usability studies), interviews, focus groups and controlled experiments (Lazar et al. 2017).

The participatory approach considers the user a partner who is included in a co-design process. This means that people who are not designers work together creatively during the design development process (Sanders and Jan Stappers 2017). The results of a UCD approach should be a well-designed system that is useful and easy to use (Preece et al. 2015). UCD helps to collect and analyse user requirements, which are the features/attributes that a product/system should include or how the product/system should perform from the users' perspective (Baxter et al. 2015). User experience is human-centred and the goal of examining the user experience is to elicit users' requirements or to evaluate an existing technology (Baxter et al. 2015). It is essential to capture the users' experiences and to understand their needs.

Various methods can be implemented to help researchers understand users, such as interviews, diary studies, surveys, card sorting, focus groups, field studies and evaluations (Baxter et al. 2015). In this paper, the focus is on exploring a range of methods: semi-structured interviews, sharing maps, diary studies, picture cards, online observation (online ethnography), co-design workshops and critique design workshops.

Qualitative interviews aim to explore and gain an in-depth understanding of a topic (Preece et al. 2015). Interviews are considered an invaluable tool in HCI research in helping the researcher to understand users' needs, practices, attitudes and behaviours (Baxter et al. 2015)(Dimond et al. 2012)(Maeng et al. 2016)(Voida et al. 2004). In particular, semi-structured interviews are conducted when researchers want to learn more about specific topics; these are based on a basic script with questions to be asked, but then probe the participants to provide more information that might be relevant (Preece et al. 2015).

There are five forms of interviews: face-to-face, video with audio, audio, interactive text (e.g. online chat) and non-interactive text (e.g. email) (Baxter et al. 2015)(Maeng et al. 2016). Researchers can choose the type most appropriate to their needs (Baxter et al. 2015). These types fall under two main categories, in person and mediated (remote) (Baxter et al. 2015)(Martin and Hanington 2012). The form of the interview can affect different aspects of the process, such as expectations of attention, timing, transcription, sharing of multimedia and conversation style (Maeng et al. 2016)(Voida et al. 2004).

Face-to-face interviews can be costly in terms of finances and time; also, there is burdens on both the interviewer and the interviewee and organising the data is a lengthy process (Maeng et al. 2016). Several studies have conducted interviews through calls (e.g. phone, Skype, etc.) to help identify practices and concepts related to the research focus or to provide an overall understanding of technology with additional, deeper insights (Al-Dawood et al. 2017)(Alghamdi et al. 2015)(Nylander and Rudström 2009). However, such modes of interviewing can also be costly and create burdens on both the participants and researcher (Maeng et al. 2016).

Other forms of interview (e.g. online messages) can be beneficial for interviewing participants who are not within driving distance of the researcher or interviewer (Hillman et al. 2015), or to include participants from different geographical regions (Alsheikh et al. 2011)(Al-Saggaf and Williamson 2004). Mediated forms are especially helpful in conducting interviews with the latter (Preece et al. 2015). Instant messages help the researcher to communicate with participants and share ideas even at a great distance (Voida et al. 2004). Participants might multitask when engaged in an interview via online messaging, e.g. sending messages to others or doing something else (Voida et al. 2004). Moreover, conducting interviews using text can save the time the researcher might otherwise have to spend on transcription (Maeng et al. 2016)(Voida et al. 2004). Therefore, mediated interviews are considered an easier option than in-person interviews (Baxter et al. 2015). However, in face-to-face interviews, there is the assurance that the participants can focus fully on the interview (Voida et al. 2004).

A previous study used quantitative analysis, focusing on word count, and qualitative codes to compare three forms of interviews, i.e. instant messaging, email and phone (Dimond et al. 2012). Although the number of words in the phone interviews was four times greater, there were no significant differences in the numbers of codes (Dimond et al. 2012). However, another study that compared phone use, instant messaging and mobile instant messaging found more data were collected using phone interviews (Maeng et al. 2016).

As already noted, interviews can either be conducted as a solo activity or in combination with another, such as card sorting or observation (Baxter et al. 2015). Integrating other methods, such as picture card sorting, can make the interviews more productive (Martin and Hanington 2012). Using picture cards is an artefact-based method, in which the researcher use images and words to help participants recall and recount actual experiences and details (Martin and Hanington 2012). Asking users in an interview how they use a system may not yield accurate answers (Lazar et al. 2017). However, giving the participants picture cards of the interfaces can gain more valid details because they help users recall their experiences and stories and they can represent their wishes (Martin and Hanington 2012). This method is more productive than others, supporting users' recall and helping them become more involved in the conversation.

The sharing map is one of the methods that can be used in interviews to understand participants' interactions with their circle with regard to online shopping (Hillman et al. 2013). A sharing map can show both the social networks participants use and – perhaps more importantly – the significant people with whom they share their shopping experiences and information.

There also methods that can be conducted alongside interviews, such as observation and diary studies (Baxter et al. 2015). Observational methods have been used, for example, as an additional means of gaining an overall understanding of interactions within the platform (Chen 2010)(Evans et al. 2018)(Moser et al. 2017). Researchers might observe the activities and interactions within an s-commerce platform to enrich the data collected (Dye et al. 2016). Observation can also help collect baseline information (Martin and Hanington 2012). Moreover, it can be used as a means of online ethnography, enabling the researcher to gather insights from online communities (Tomitsch et al. 2018). Online ethnography can be conducted to investigate issues related to the online community (Gheitasy et al. 2015). This method helps to study online interactions among those participating in an online platform (Tomitsch et al. 2018).

Diary studies can be used to collect data across time, including participants' thoughts, feelings and behaviours over a specific period (Martin and Hanington 2012). Diary studies are usually conducted with other methods, such as interviews (Baxter et al. 2015). Moreover, digital diaries can be used, enabling participants to submit their diaries online and include photos and text entries (Martin and Hanington 2012). The benefits of this method are that it can help researchers enrich the data collected without the need to be present (Baxter et al. 2015).

The other approach is participatory design (PD), which can be defined as "a set of theories, practices, and studies related to end-users as full participants in activities leading to software and hardware computer products and computer-based activities" (Muller 2012). PD can also be called co-creation/co-design (Sanders and Jan Stappers 2017). In PD methods, all the participants contribute to discussions during workshops (Dillahunt and Malone 2015). Co-design has been growing in the PD field (Sanders and Jan Stappers 2017); co-design workshops help to ensure that the users' needs are addressed during the design process, the application being designed with them rather than for them (Tomitsch et al. 2018). The co-design process can help provide users with a higher quality service experience that better meets their needs (Steen et al. 2011). Co-design workshops can be used in any stage of the design process (Tomitsch et al. 2018). They can help designers to generate better ideas and acquire greater knowledge about customers' needs (Steen et al. 2011). They can be used in the research phase to help gain a complete overview of the situation (Tomitsch et al. 2018). They can also be used during the prototype phase to iterate concepts rapidly (Tomitsch et al. 2018).

An early method for evaluate design is to conduct design critique workshops, which can help to reflect on the design rather than testing it (Greenberg and Buxton 2008). Design critique focuses on evaluating existing ideas, not developing new ideas (Tomitsch et al. 2018). Such evaluation can be undertaken by presenting the design to the participants and asking them to identify what they like and dislike (Greenberg and Buxton 2008)(Tomitsch et al. 2018).

## Qualitative and design studies in social (s-)commerce

The majority of existing studies in s-commerce have been quantitative and have employed surveys (Esmaeili and Hashemi G 2019)(Han et al. 2018)(Zhang and Benyoucef 2016). It is recommended that future studies in s-commerce conduct qualitative research to obtain more empirical evidence with respect to consumer behaviour (Han et al. 2018) as few studies have used qualitative methods to understand users' behaviours and needs with regard to s-commerce (Moser et al. 2017)(Shin 2013).

A previous study explored the use of Facebook as an s-commerce platform to trade products within a mom-to-mom buy and sell group (Moser et al. 2017), with the mothers as both buyers and sellers. The study entailed semi-structured interviews and observation of the group for five months to collect field notes of the members' interactions. Observation helped contribute to a general understanding of how Facebook buy and sell groups work. Another study investigated the use of Facebook to conduct transactions (Evans et al. 2018). The methods used were semi-structured interviews and monitoring of Facebook bartering groups for eight months, which provided an in-depth understanding of the interactions between members. Another study was also conducted with Facebook to identify the factors that affect consumers' intentions (Marjan et al. 2014). Interviews were undertaken to identify new factors that can affect consumers' intentions and the reasons why.

In a study that used both semi-structured interviews and sharing maps to capture personal data on shopping and relations (Hillman et al. 2013), the use of group shopping and friendship networks was investigated to propose the optimal design for s-commerce based on the consumers' experiences with a view to improving the user experience. Another study on s-commerce examined the impact of social information on consumers' decision processes (Chen 2010). Semi-structured interviews were conducted after the participants performed a task using two sites. The data provided in-depth information regarding the motivation behind consumers' choice of product. Moreover, observing the participants while they performed the tasks provided information concerning social features and their role in buying a product. Furthermore, a previous study explored the role of s-commerce in enhancing consumers' trust (Ji et al. 2019). Again, semi-structured interviews were conducted, as well as direct observation of the s-commerce platforms (companies' websites and WeChat). These methods helped identify issues related to consumers' trust.

A previous study that conducted mock-ups used a survey as the main research method (Suraworachet et al. 2012). The survey employed mock-ups, incorporating a scenario and a rating scale, then asked for any additional suggestions. It evaluated the effect of s-commerce features on intention to buy over Facebook. Another study used experiment-based surveys to explore the influence of s-commerce features on users' attitudes towards a manufactured e-commerce website (Friedrich et al. 2016). The e-commerce website was created based on previous literature. The researchers provided design implications based on their results.

Although these studies investigated the use of s-commerce and consumers' behaviours, they did not include sufficient details about the role of the features and design in enhancing consumers' trust. Moreover, most of the studies focused on Facebook, WeChat and various websites. As far as can be ascertained, there are no extant studies on s-commerce that have used co-design or critique workshops. There are only limited studies that have discussed the design of s-commerce platforms, the features necessary and their impact.

## Sample size

The recommended sample size for interviews examining users' perceptions varies. One study recommended having at least six participants to gather users' perceptions (Huang et al. 2012). However, based on expert opinion, another argued that the optimal number of participants depends on the method, i.e. 12–20 participants for interviews and 4–12 participants per group for focus groups, with 3–4 groups in all (Baxter et al. 2015). Previous studies that have been conducted to understand users' experiences and report their perceptions have recruited 14–18 participants (Al-Dawood et al. 2017)(Al-Dawood et al. 2020)(Al-Saggaf and Williamson 2004)(Hillman et al. 2013)(Nassir and Leong 2018). For the co-design workshops and design critique workshops in this study, it was recommended to have 3–10 participants (Tomitsch et al. 2018).

## Culture

The main challenges that may face researchers in Saudi Arabia are participants' privacy concerns when talking to a stranger and cross-gender communication (Abokhodair and Vieweg 2016)(Nassir and Leong 2017). It is difficult to recruit research participants in Saudi Arabia (Abokhodair and Vieweg 2016), especially due to privacy concerns, which can be significant (Nassir and Leong 2017). To make it easier to secure a sample, the snowball sampling approach should be used, as recommended by others [1], to help participants trust the researcher to a certain extent (Nassir and Leong 2017). Using this means, the researcher recruits participants who are recommended by personal and professional connections and they in turn recruit others (Abokhodair and Vieweg 2016). However, this approach may not result in participants with diverse backgrounds (Al-Dawood et al. 2017)(Al-Dawood et al. 2020). Therefore, it is useful also to use another approach, for example by sending invitations through social networks (Alghamdi et al. 2015) (including an online sign-up form (Al-Dawood et al. 2020)), mixing approaches to reach a wider range of participants from diverse backgrounds (Al-Dawood et al. 2017)(Al-Dawood et al. 2020). Recruiting participants by email has been found not to be as effective as instant messaging (Al-Saggaf and Williamson 2004). It is also important to describe the data collection method(s) in the invitation letter, particularly as some methods, such as focus groups, are not commonly used in Saudi Arabia (Alghamdi et al. 2015). Indeed, although Saudi Arabia is a modern society, it is conservative (Al-Saggaf 2011) and therefore great consideration should be given to the data collection methods used when conducting studies in studies, such as interviews (Al-Dawood et al. 2017)(Al-Saggaf 2011).

In cross-gender communication, entailing interaction between females and males, the women should be chaperoned (Alsheikh et al. 2011). Moreover, a male researcher may ask female participants to gain permission from their guardians prior to conducting interviews (Al-Dawood et al. 2017)(Al-Saggaf 2011). It has been considered essential not to initiate direct contact with females, but to draw on male relatives to mediate talks (Nassir and Leong 2018). However, the law of guardianship changed in 2019 to allow women above 21 years of age to travel without permission and to apply for a passport (Krimly 2020). Therefore, asking the guardian's permission might not be necessary from the legal perspective, but the cultural norms persist (Al-Dawood et al. 2020).

For chaperoned interviews, the researcher can use probes to collect data supplementing the initial responses as the answers from participants might be affected by the presence of the chaperones (Nassir and Leong 2018). Moreover, online communication can be beneficial for several reasons. First, it facilitates cross-gender communication (Al-Saggaf and Williamson 2004). As well as being low cost, online communications have been found to be effective in gaining in-depth information about an individual's experiences (Al-Saggaf and Williamson 2004). Sometimes, female participants might prefer to be interviewed anonymously via online chat, for example using Facebook Messenger, to prevent communicating with a (male) stranger (Al-Dawood et al. 2020)(Almakrami 2015). Conducting interviews by telephone is also considered an appropriate method due to the culture of gender segregation (Alghamdi et al. 2015). Female participants can also communicate with a male researcher and pass on information through social media (Nassir and Leong 2018). Equally, it is better for a female researcher to conduct interviews with male participants by telephone, not face to face (Flechais et al. 2013), or through social networks to avoid the need for the researcher to have a chaperone (Nassir and Leong 2018).

For qualitative studies in Saudi Arabia, social media can be a good source of information from participants, especially in the case of cross-gender communication (Nassir and Leong 2018). Participants may be willing to share photographs and videos with researchers using social networking (Nassir and Leong 2018). However, conducting an initial interview before asking participants to share information over social media may help to build trust, especially as the participants know more about the researcher and vice-versa (Nassir and Leong 2018).

Deciding where to conduct a face-to-face meeting is another challenge that faces the researcher as it is not permitted to do so in private premises and rent a meeting room is an expensive option (Alghamdi et al. 2015). It is common in Saudi Arabia for females and males to have separate spaces in places such as schools, universities and government departments (Flechais et al. 2013). Therefore, separate sessions for focus groups should be conducted for females and males in different gender-appropriate locations, the female sessions being run by a female researcher/moderator and the male sessions being run by a male researcher/moderator (Alghamdi et al. 2015).

Interviews should be conducted in a public place or workplace (Flechais et al. 2013). A public place could be a restaurant that provides family rooms (Alghamdi et al. 2015). While interviews can be undertaken in the participant's home, it can be a challenging environment because of the presence of other family members and the potential for interruptions and noise, for example from children and extended family (Nassir and Leong 2018). It is not usual to interview participants in their homes even if they are the same gender (Alghamdi et al. 2015), but a previous study conducted in the western region of Saudi Arabia, where gender segregation is more relaxed, did so (Nassir and Leong 2018). The researcher, who recruited participants through snowball sampling, was treated as a guest and provided with hospitality (Nassir and Leong 2018).

In Saudi Arabia, many participants may reject compensation for participation, for example a gift card, preferring to take part voluntarily (Al-Dawood et al. 2017). It might be more appropriate to give a gift, such as chocolates or a pen, rather than cash or a gift card (Al-Dawood et al. 2020)(Nassir and Leong 2017). Indeed, compensation for Saudi participants should not be monetary.

## The context of Saudi Arabia

In Saudi Arabia, s-commerce has grown rapidly (Alotaibi et al. 2019). It has been used by women entrepreneurs to present and sell their products, for example, those that are handmade or customised. Previous studies have examined entrepreneurs' use of social media applications to conduct commercial activities, such as implementing a marketing strategy (Alotaibi et al. 2019)(Alghamdi and Reilly 2013). Such studies used questionnaires or online observation and recorded social media data through Web analytics tools. They found that there are cultural boundaries (i.e. familial, governmental and political). However, their results are still being developed. Other studies have focused on factors that are positively associated with behavioural intentions, such as social support and s-commerce constructs (Makki and Chang 2015)(Sheikh et al. 2017), as well as factors that influence the adoption of s-commerce (Abed et al. 2015). They all employed a quantitative approach based on questionnaires.

There are very few studies that have investigated factors affecting consumer trust. Examples of such studies include that of Alotaibi, Alkhathlan, and Alzeer (Alotaibi et al. 2019), which looks at factors such as social media influencers (SMIs), key opinion leaders (KOLs), and consumer feedback. Another interesting work demonstrates how trust mediates the relationship between social support and s-commerce intentions (Al-Tit et al. 2020). However, in general, there is a lack of research focused on identifying the mechanisms that can enhance trust in online s-commerce through design. Moreover, the studies that have been conducted only tested a limited number of factors and solely employed a quantitative approach.

# Methods

This study employed a user-centred and participatory approach to understand the use of s-commerce and how its features influence users. The research comprised five studies aimed at understanding user behaviour and designing and evaluating an s-commerce platform. The main aim was to understand the current use of s-commerce and identify how the user experience and enhance trust focusing on the features of the application. Table 1 shows the methods and tools used, as well as the number of participants in each study.

The first study entailed semi-structured interviews with buyers conducted by telephone and instant messaging (e.g. WhatsApp). The questions asked about their use of social media applications, the commercial use of social media applications and the use of e-commerce websites.

The second study comprised semi-structured, face-to-face interviews with buyers. The interview protocol had three parts: (i) asking participants to draw a sharing map of their social networks and the people with whom they shared their shopping information; (ii) asking participants to imagine conducting commercial activity through a social network and through an e-commerce website and explain the differences in use and issues of trust; (iii) showing picture cards and asking participants what features might help them make a purchase decision.

The third study conducted semi-structured interviews with sellers to identify how they gained customers' trust. The interviews were either conducted face to face or through instant messaging. Online observation was also undertaken to examine activities and interactions.

The fourth study comprised a co-design workshop to inform on the design of an s-commerce platform and identify the features that should be included. The workshop included screenshots of Instagram and Amazon to help participants describe what they liked on these platforms and how one might be better designed. Participants were asked to sketch a prototype, but they were unlikely to do so. Therefore, they were given cards showing the features/functions and their icons and asked to sort these cards into three groups according to importance: very important, important and not important.

Finally, the fifth study entailed the design critique workshops, presenting mock-ups including the very important and important features and asking the participants to critique the design. In these workshops, participants were asked to identify what they liked and what they did not understand and their further suggestions for the design.

Table 1 Summary of methods conducted in the study

|   | Main method(s) | Tools | Number of participants |
|---|---|---|---|
| 1 | Semi-structured interviews with buyers | By text | 18 |
|   |   | By voice (telephone or online such as Skype) | 8 |
| 2 |   | Face to face with activities | 17 |
| 3 | Semi-structured interviews and online observation with sellers | Face to face | 7 |
| | | By text | 10 |
| 4 | Co-design workshops | Screenshots of interfaces (Instagram and Amazon) + online access to platforms | 10 |
| 5 | Critique workshops | System usability scale (SUS) + interactive prototype + screenshots of the interactive prototype | 16 |

## Recruitment

Three methods were used to recruit participants: sending emails, posting an invitation on social networks and snowball sampling. Posting the invitation on social networks and the snowball method were more effective than the email. A Google form was created and included in the invitation with an information sheet and description of the study. For all the studies conducted, participants were asked if they could/would take part in interviews or workshops online.

For the co-design and critique design workshops, the invitation included a message asking those happy to participate to click on the link to the Google form. The invitation also stated that I, as the researcher, would be happy to answer any questions before participants registered. For the workshops, few participants registered initially. I received several private messages asking almost identical questions about the workshop: where would it be, could they attend virtually, was it mixed gender or only female, did they need to draw something or just talk? After answering these questions directly, some of them registered. I also resent the invitation through social networks, addressing these questions in the invitation as most would not open the information or the link before knowing the basic details.

## Results

The following sub-sections discuss various aspects concerning the implementation of the methods used in the different studies.

## Time

The telephone interviews lasted, on average, 15 minutes. The face-to-face interviews took around 20–40 minutes. The face-to-face interviews that also included activities naturally took longer than those that did not. However, the online chats took longer, especially as sometimes the participants were busy and took 1 or 2 hours to respond, for which they apologised. Although the online interactions took longer, it was not necessary to wait to reply to participants, as I could respond immediately when I received a notification.

The face-to-face interviews were more time consuming. Travelling to the interview location took around 30 minutes as the routes in Saudi Arabia are increasingly crowded. Moreover, on arriving at the location, some of the participants did not attend, apologising that something urgent had come up. With the online messages and telephone interviews, there was no need to spend time on travelling to the location. Also, online interviews could be conducted by other means, such as mobile telephones, and they could take place at any time and in any place. Some participants interviewed through online chat sent voice notes in response instead of typing their answers, which was even better because they included more details. However, many participants preferred to type rather than sending voice messages.

In terms of the workshops, some participants who were registered did not attend, sending apologies that they had to stay with their children or they did not feel well. The workshops had to start on time as the other participants might have later appointments. To overcome the issues with last minute cancellations, I recruited the maximum number of potential participants who might attend the workshop, so that even if a few did not attend, the overall participation was reasonable.

## Response: Concentration and engagement

The response, level of attention and engagement from users differed based on the form of the interview and location. Interviewing by telephone presented some challenges. First, some participants did not live in a quiet location, making it difficult to hear them or to have a smooth conversation. Some were at home with their family and children, which resulted in background noise and the participants being distracted by family members asking questions. Sometimes, the participants received another call during the interview and asked if they could just take it quickly, which was time consuming. In some cases, the participants were walking or driving during the interviews, which could be noisy and meant they were not entirely focused on the interview.

In contrast, interviewing through online messaging, for example using WhatsApp, Skype, Line, etc., helped ensure a smooth conversation without any noise. However, some participants would multitask, responding to other messages or doing something else. Before the interviews started, some participants asked if it would be alright to be "in and out", saying they would prefer this as they would be able to concentrate and answer the questions fully if, for example, they could leave and then return to a question after dealing with an interruption from a family member or background noise.

Regarding the face-to-face interviews, the attention of the participants depended on the location. The interviews held in a public place, such as coffee houses, were sometimes disrupted by noise when the café was busy. Other interviews were conducted in a meeting room at the participant's workplace or in the participant's office. In the former instance, the participants were able to concentrated fully on the interview, but in the latter case they were sometimes interrupted by colleagues calling on the telephone or dropping by to ask about something.

In terms of engagement, each form differed. In the online chat mode, it was easy to share multimedia and to ask questions freely and be open to answering questions. The participants also shared pictures of their experiences or photographs of the applications that they liked. However, with the telephone interview, it was not possible to share any images and the participants did not really ask questions; they just answered my questions and ended the interview as soon as possible. In the face-to-face interviews, I could show the participants picture cards and we could discuss them, as in online messaging. However, there were also other activities that encouraged the participants to discuss their experiences in greater depth, such as drawing a map of their social media application use and accessing their mobile devices and imagining ordering through s-commerce/ e-commerce and how they would go about it.

In terms of online chat, it was easier to follow up with participants and ask them if additional information was needed to verify their meaning. Also, the participants shared screenshots and photographs of their shopping experiences and sent them through WhatsApp with comments (e.g. that they knew the shop account from their friends or a screenshot of a recommendation from their friends or family).

In the co-design workshops, some features were not identified as important by users, but when they saw them in the mock-ups, they thought them really helpful and important. For example, the reviewer's profile was not considered important in the co-design workshops, but in the critique design workshop, they checked the profile to ensure it was not fake and viewed it as helpful.

## Culture

There were some challenges that might be related to culture, for instance in terms of cross-gender communications, geographical issues (distance), time (cancellation), participants' concerns about privacy, hesitation from some participants and issues of compensation.

It was difficult to have face-to-face interviews across the genders. Therefore, male participants were asked to choose if they wanted to conduct interviews by telephone or through online chat. The workshops comprised only female participants as it was not possible for them to be mixed gender or to ask someone else to conduct sessions for male participants.

In terms of distance and geographic challenges, conducting interviews in different locations in Saudi Arabia could be time consuming and costly, requiring travel between locations sometimes lasting more than 1 hour. Also, sometimes the interviewee would be late or not arrive at all. Therefore, online chat interviews were more convenient. For the workshops, the potential for non-attendance, which did happen, was taken into account by inviting larger numbers than required.

Consistent with research guidelines (e.g. Code of Good Practice in Research and Handling of Personal and Sensitive Data guidelines from Newcastle university), the participants were informed that their information and details would be totally anonymised. Also, they were told that in this study they were the experts in terms of how to use the platforms, helping give them the confidence to explain even the simplest details or activities. In some cases, the participants asked if they had provided the "right" answer and therefore they were informed before the interviews that there were no correct or wrong answers or correct or wrong behaviours and that the answers only aimed to understand the use of s-commerce.

For the design critique workshops, the participants were informed that the design had been worked on by previous participants and they were free to critique it. This helped the participants feel free to critique the design.

Finally, as already noted, many participants rejected monetary compensation in the form of a gift card. They kept saying that they wanted to participate on a voluntary basis. In some cases, I therefore gave them chocolate, which was more acceptable.

## Transcription and analysis

Transcribing interviews and workshops takes time and there is no readily available software that can do it effectively for Arabic. Some websites claim to provide auto-transcription in Arabic, but it does not match at all. Therefore, transcription was undertaken manually, which was very time-consuming: for each hour of recording, it might take three to four hours to transcribe, especially with background noise. Interviews undertaken in online chat were effectively already transcribed, being in text, which meant less effort. However, the interactions in the workshops were more difficult to transcribe as there were multiple people and sometimes two groups were talking at the same time. To make it possible to distinguish what the two groups were saying, individual recording devices were used for each.

Conducting online chat interviews would potentially save time on transcription as the interviews would already be in text form (unless using voice notes). In contrast, the telephone and face-to-face interviews needed to be transcribed. The telephone interviews were more difficult to transcribe due to background noise, both on the participants' end and speaker interference. The face-to-face interviews were easier to transcribe as we were both (interviewer and interviewee) in the same location and there was minimal or even no background noise, especially when the interviews were conducted in meeting rooms or offices.

The language used in this study was Arabic and ATLAS.ti was used to help analyse the data as the program supports Arabic. In contrast, NVivo, often used for qualitative analysis, was not suitable for use with Arabic script and several issues arose when trying to employ it, such as the imported file not showing all the text and it skipping when trying to code a phrase.

## Qualitative coding and word count

Word counts and codes were tested to check if there were significant differences between different data collection methods and forums. A global word count was used without deleting repetition. Four comparisons were made as follows: (i) online chat and telephone; (ii) online chat and face to face; (iii) face to face, online chat and telephone; (iv) co-design workshops and design critique workshops.

- **Comparison of online chat and telephone interviews (Study 1)**

Table 2 Median scores for online chat and telephone interview data

| Interview form | Word count | Codes |
| --- | --- | --- |
| **Online chat** | 1426 | 43 |
| **Telephone** | 741 | 49 |

As can be seen from Table 2, although the word count for online chat is greater, the number of codes for telephone is higher. While the difference in the word count is significant ($p = 0.001$), the difference in the number of codes is not ($p = 0.101$) and therefore, there are no significant differences in the information obtained from the two methods.

- **Comparison of online chat and face-to-face interviews (Study 3)**

Table 3 Median scores for online chat and face-to-face interviews

| Interview form | Word count | Codes |
| --- | --- | --- |
| **Online chat** | 3297 | 111 |
| **Face to face** | 2146 | 85 |

As is apparent in Table 3, the word count and number of codes were both higher in the online chat interviews. The differences in both instances are significant (p = 0.001). This result suggests that online chat can obtain more information than face-to-face interviews. It should be noted that in this case, the interviews entailed asking and answering questions and there were no additional activities.

- **Comparison of online chat, telephone and face-to-face interviews including activities (Studies 1 and 2)**

Table 4 Median scores for online chat, telephone and face-to-face interviews with activities

| Interview form | Word count | Codes |
| --- | --- | --- |
| Online chat | 1426 | 43 |
| Telephone | 741 | 49 |
| Face to face | 1868 | 60 |

Table 4 shows that the word count for the face-to-face interviews (including activities) and the number of codes are higher than for the telephone and online chat interviews. Moreover, the difference in word count between face-to-face interviews and online chat is significant (p = 0.001) and the difference in the number of codes is between face-to-face interviews and telephone interviews is also significant (p = 0.005). Thus, it appears that face-to-face interviews including activities can provide richer data and greater detail than other methods of interviewing.

- **Comparison of co-design and critique design workshops**

Table 5 Median scores for co-design workshops and critique design workshops

| **Interview form** | **Word count** | **Codes** |
| --- | --- | --- |
| **Co-design workshop** | 5250 | 290 |
| **Critique design workshop** | 3681 | 149 |

As shown in Table 5, the co-design workshops present both a higher word count and a greater number of codes than the critique design workshops, a difference that is significant (p = 0.001). This is most likely because the co-design workshops included more discussion of previous experiences and stories of shopping issues that participants might face. In the critique design workshops, the participants were focused on evaluating the design in terms of what they liked/disliked and suggestions.

# Discussion and conclusion

Semi-structured interviews including no additional activities can be undertaken through online chat, by telephone or face to face. Based on the results, online chat and telephone are better at eliciting information than face-to-face interviews. However, if semi-structured interviews include activities, participants can be encouraged to provide more information.

The word count is not particularly important as sometimes the participants repeat what they have said or amplify rather than just giving direct answers. In contrast, the number of codes potentially reflects the provision of greater information.

Online chat interviews are the cheapest option and require less effort on the part of the researcher as they can be undertaken anytime and anywhere, there are no issues with cross-gender communication, the participants and researcher are not stressed and participants are comfortable about saying whatever they want. It is also beneficial to use online chat as participants can share photographs, providing rich data and additional details.

Telephone interviews can be costly in terms of time and require total focus. Not only is it necessary to address potential noise and signal problems, but also participants may be busy or distracted and cut questions, which is distracting. Recordings can be unclear and they can be difficult to transcribe. On the other hand, they can provide rich data.

Face-to-face interviews not including activities are the least recommended means of collecting data as they require considerable effort. They can be costly because of having to travel to the meeting place, waiting for participants to show up and transcribing the interviews. The data gathered could be less informative than those obtained through telephone interviews and online chat. However, if the interviews include activities, such as sharing maps, using mobile phones to recall shopping experiences, or using picture cards to discuss the features, face-to-face interviews can provide richer data and more details than telephone interviews or online chat. It is important to bear in mind the cost, time and effort in selecting the optimal method.

Finally, co-design and critique design workshops can be beneficial in informing on the optimal design. The co-design workshops included a wealth of information as the participants explained their previous experiences. However, in the critique design workshops, the participants were focused more on the existing design and commenting on aspects related solely to the design. It is necessary to employ a mix of methods to learn about users' needs and experiences. Also, it is useful to incorporate more than one method to validate previous results (Steen et al. 2011). Table 6 summarises the advantages and disadvantages of the methods implemented in this research.

Table 6 Summary of advantages and disadvantages of methods conducted in this research

| Method | Advantages | Disadvantages |
|---|---|---|
| Semi-structured interviews – by text | <ul><li>Preferred by participants as they can take place while they are at home with their family.</li><li>There is no need to travel</li><li>Can be done over wide geographic distances</li><li>More comfortable for cross-gender communication</li><li>Can keep going without the feeling that they want to end it</li><li>No need to transcribe</li></ul> | <ul><li>Can be slower than other methods</li><li>May take a longer time to finish the interview</li></ul> |
| Semi-structured interviews – by voice | <ul><li>More comfortable for cross-gender communication</li><li>Preferred by participants as they do not need to travel and meet in a specific place</li><li>Do not need a location</li><li>Saves researcher's time (travelling to the location and waiting for the participants)</li></ul> | <ul><li>Could be unclear (e.g. signalling issues)</li><li>Could have background noise (e.g. children, family members, car, people)</li><li>Might want to end the call sooner than desirable</li><li>Hard to transcribe</li></ul> |
| Semi-structured interviews – face to face | <ul><li>Interviewee focus is mostly on the interview (depending on the location)</li></ul> | <ul><li>Needs a location</li><li>Takes time to travel to the location and wait for the participants</li><li>Takes time to transcribe</li></ul> |
| Online observation | <ul><li>Helps to provide overall information about the general interaction and activities</li><li>Helps to support the data collected by other methods such as interviews</li></ul> | <ul><li>If the platform does not make it possible to collect data through a tool, it is more difficult to conduct observation</li><li>It is not possible to observe private interactions and activities</li><li>It can be time-consuming as the observation needs to take place over a period of time (several hours at a time over several weeks/months)</li></ul> |
| Semi-structured interviews – face to face with picture cards | <ul><li>Helps participants provide more details</li><li>Helps participants remember the details of previous experiences</li><li>Participants are more open to talking as they look at the drawing and remember to mention who they share experiences with</li><li>Can help make interviews interesting and keep the participant engaged</li></ul> | <ul><li>Time-consuming to travel to the location and to transcribe</li><li>Need for a location</li><li>Difficult to conduct cross-gender interviews</li></ul> |
| Diary | <ul><li>Provides additional information (experiences) to supplement interviews</li></ul> | <ul><li>Participants forget to share diary entries and the researcher needs to send reminders</li></ul> |
| Co-design workshops | <ul><li>Helps to understand needs and issues</li><li>Having more than one participant can bring new ideas and insights to the discussion</li><li>Discussing previous experiences and challenges</li><li>Using card sorting helps participants discuss issues in greater depth and show what they want and need</li></ul> | <ul><li>Some participants might not wish to share thoughts, so the researcher should try to keep the discussion going and include all the participants</li><li>Some participants might start to talk about things not related to the workshop; then the researcher should again bring the discussion back on topic</li><li>Participants might not want to sketch or draw</li></ul> |
| Critique workshops | <ul><li>Participants will focus on critiquing the design and specify what they like</li><li>Can help to improve the design</li><li>Helps to check how they interact with the system and they might realise the need for features not previously considered before interacting with the platform</li></ul> | <ul><li>Participants might start to focus on specific details, such as colours or photos</li><li>Some participants might start to talk about things not related to the workshop; then the researcher should again bring the discussion back on topic</li><li>Some participants might not share their thoughts, so the researcher should try to keep the discussion going and include all the participants</li></ul> |
| SUS | <ul><li>Helps identify if there is general acceptance from the users</li><li>Identifies the extent to which the system is easy to use and understandable</li></ul> | <ul><li>Might not be accurate in testing an interactive prototype as not all the functions can be used</li><li>Not a useful method to be used alone as the researcher needs to know the difficulties that the participants face or issues with the system</li></ul> |

- **Brief recommendations for conducting the methods discussed**

Before implementation, it is important to explain to the participants that they are free to say anything related to the topic and that there are no correct or wrong answers. Also, the researcher should make it clear to the participants that they are the experts and the researcher wants to learn from and understand their experiences. Finally, it is recommended that Saudi participants are given a choice between monetary compensation (e.g. a gift card) or a gift (e.g. a small box of chocolates).

**For online interviews**
- Prepare the main questions in a note, making it possible to copy and paste question rather than typing each while the participants wait.
- Decide with the participants before the interview whether it is alright to be "in and out" or whether it should be done in one go at the same time.
- Inform the participants that they are free to send voice notes rather than typing their answers.
- Prepare some pictures to share with participants as prompts, encouraging them to talk more extensively.
- Inform the participants that if they remember anything and they want to add it later, they can send information at any time, thus providing as much as relevant information as possible.

**For telephone interviews**
- Ask the participants before the interview to schedule a time that they can be in a quiet place and not busy.
- Ask the participants to send any additional information they want through online chat.

**For face-to-face interviews**
- Prepare activities to undertake during the interviews rather than just asking questions and waiting for answers.
- Choose a quiet location to conduct the interviews.

**For workshops**
- Be sure that all participants are participating in the discussion.
- Be sure that the discussion is related to the research focus and questions.
- Divide the workshops into different activities.
- Have a break in the middle of the workshops and provide some refreshments (e.g. chocolate and coffee).

# Limitations and future work

Despite the different methods employed to understand the participants' needs for certain features and use of s-commerce, there are some limitations to this research. First, the study was conducted in Saudi Arabia and all participants were Saudi. For future research, the same methods could be conducted in different cultures, comparing the results. Second, some of the methods (face-to-face interviews and workshops) were only conducted with female participants due to the culture. In future, these methods could also be conducted with male participants for comparison. Moreover, cross-gender face-to-face interviews could be undertaken to identify differences in perceptions and determine the feasibility of conducting such research after the changes in Saudi Arabia. Finally, future research could compare methods other than those used in this study.